\documentclass[universe,article,accept,oneauthor,latex]{mdpi}     
\firstpage{1} 
\pubvolume{1}
\issuenum{1}
\articlenumber{0}
\pubyear{2022}
\copyrightyear{2022}
\externaleditor{}

\datereceived{} 
\dateaccepted{} 
\datepublished{} 
\hreflink{} 

\Title{Selected Results in Heavy-Quark Fragmentation}

\TitleCitation{Selected Results in Heavy-Quark Fragmentation}

\Author{Gennaro
 Corcella  \orcidA{}}

\AuthorNames{Gennaro Corcella}

\AuthorCitation{Corcella, G. }

\address[1]{ INFN, Laboratori Nazionali di Frascati, Via E. Fermi 54,
  00044 Frascati, RM, Italy; gennaro.corcella@lnf.infn.it}

\abstract{I review a few selected topics concerning heavy-quark fragmentation, taking particular care about bottom- and charm-quark
  production in $e^+e^-$ annihilation and the inclusion of non-perturbative
  corrections.
  In particular, I discuss the recent developments of calculations carried
  out in the framework of perturbative fragmentation functions and the
  perspective to extend them to other processes and higher accuracy.
  Special attention is
  paid to the use of an effective strong coupling constant to model
hadronization effects.}


\begin{document}

\section{Introduction}
Heavy-quark phenomenology (top, bottom, and charm)
is one of the most interesting research topics in
particle physics, since it allows tests of the Standard Model in both
strong and electroweak sectors and often plays a role in several searches for
new physics phenomena.
As for QCD, on which the present contribution will be mostly concentrated,
processes with heavy quarks offer the opportunity to test the parton model,
factorization, and power corrections. 

In principle, as the mass of the heavy quarks regulates the collinear
singularity, one should trust perturbative QCD to describe heavy-quark
production. However, it turns out that fixed-order
calculations, e.g., at next- (NLO) or next-to-next-to-leading order
(NNLO) in the strong coupling constant $\alpha_S$, are reliable to predict
inclusive observables, namely total cross-sections or widths, while differential
distributions, such as the energy or transverse momentum spectrum of the
heavy quark, exhibit large logarithms which need to be resummed to
all orders to obtain meaningful results.

For this purpose, the approach of perturbative fragmentation functions
\cite{mele} stands out as a powerful tool. Regarding up to power
corrections $\sim (m/Q)^p$, with $m$ being the heavy-quark mass and $Q$ the
process hard scale, e.g., the centre-of-mass
energy in $e^+e^-$ annihilation, the heavy-quark energy spectrum
can be obtained as a convolution of a massless coefficient function and a
process-independent
perturbative fragmentation function, which describes the transition of the
massless parton into the heavy quark. This factorizazion, as long as
$m\ll Q$, allows one to predict the spectrum of a massive quark just by
performing a massless computation.
Heavy-hadron energy distributions are then determined by convoluting
the parton-level ones with hadronization models which contain
some tunable parameters, such as, e.g.,
the non-perturbative fragmentation functions in \cite{kart,pet}
\endnote{See also \cite{shif1,shif2} for pioneering work on
non-perturbative effects in Quantum Chromodynamics.}.

This approach has been applied to bottom/charm  production
in $e^+e^-$ collisions \cite{mele,cc,cno}, $b$-quark
fragmentation in top \cite{cm,ccm} and Higgs \cite{corc}
decays, and charm production in charged-current Deep Inelastic Scattering
\cite{cmdis}. 
All these papers compute the energy spectrum at NLO and evolve
the perturbative fragmentation function from the hard scale down to the
heavy-quark mass by using the
Dokshitzer--Gribov--Lipatov--Altarelli--Parisi (DGLAP) evolution equations
\cite{alta,dok,gribov}. One can easily demonstrate that,
once the initial condition of the perturbative fragmentation function
is given, solving the
DGLAP equations allows one to resum large logarithms $\ln(m^2/Q^2)$
(collinear resummation). In particular,
in these papers, collinear resummation was carried out in the
next-to-leading logarithmic (NLL) approximation.
The initial condition was computed at NLO in \cite{mele} and its
process independence was proved on general grounds in \cite{cc}.
More recently, it was calculated in \cite{melni} and \cite{alex}
at NNLO for heavy quarks and gluons, respectively.
Such results, along with the NNLO computation of the Altarelli--Parisi
splitting functions in \cite{mmv}, would permit DGLAP evolution and collinear
resummation up to NNLL.
Furthermore,
the NLO initial condition of the perturbative fragmentation function
and the coefficient functions of processes \cite{mele,cno,corc} exhibit
terms which are enhanced whenever the heavy-quark 
energy fraction becomes close to 1, which corresponds to soft or collinear
emissions. Such terms can be resummed using standard techniques
\cite{sterman,catani}, achieving what is called large-$x$, soft, or threshold
resummation. The authors of ~\cite{cc,cno,ccm,corc} accounted for
threshold resummation in the NLL approximation in both coefficient functions and
initial conditions.

Regarding the inclusion of hadronization effects, an alternative approach
to the use of a model with tunable parameters
involves incorporating non-perturbative corrections into a
frozen \cite{dokweb} or analytic \cite{shirkov} strong coupling constant.
For the purpose of heavy-quark fragmentation, the model in \cite{shirkov}
was used in \cite{agl,corfer} to predict $B$- and $D$-hadron
production in electron--positron annihilation, in conjunction with
a NLO coefficient function, NLL DGLAP evolution, and NNLL large-$x$
resummation.

Before concluding this section, I wish to point out that, although
the present manuscript will not deal with heavy-quarkonium production,
relevant work on the fragmentation to heavy quarkonium can be found, e.g., in
~\cite{braat1,braat2,lei}.

The present paper is structured as follows.
In Section \ref{secper}, the approach of perturbative fragmentation will be reviewed.
In Section \ref{sec3},  the implementation of hadronization effects will
be discussed, while a few phenomenological results will be presented
in \mbox{Sections \ref{sec4}. Finally, Section \ref{sec5}} will contain some concluding remarks.

\section{Perturbative Calculations for Heavy-Quark
Fragmentation}\label{secper}

As discussed in the introduction, the framework of perturbative fragmentation
functions represents a powerful tool to compute the heavy-quark energy
spectrum:
as long as the quark mass is negligible with respect to the hard scale,
the heavy-quark energy distribution
can be obtained through a convolution of a massless coefficient function and
a process-independent perturbative fragmentation function.
Concerning the computation of the energy spectrum in the massless
approximation, 
it is well known that it is divergent
because of the collinear singularity, which has to be subtracted in order to
obtain a finite result and consistently define the coefficient function.
At NLO, the calculation is typically carried out in dimensional regularization,
with the collinear singularity subtracted off in 
the $\overline{\rm {MS}}$ scheme. \endnote{After performing a calculation
  in $d=4-2\epsilon$ dimensions, the  $\overline{\rm {MS}}$
  scheme is defined in such a way that the term $\sim(-1/\epsilon
  +\gamma_E-\ln(4\pi))$ is subtracted off.}
Referring to $e^+e^-$ annihilation into
bottom-quark pairs at the $Z$ pole at ${\cal O}(\alpha_S)$ for simplicity, i.e.,
\begin{equation}
  e^+e^-\to Z(Q)\to b(p_b) \bar b(p_{\bar b}) (g(p_g)),
  \end{equation}
the energy distribution can be factorized
as follows:
\begin{eqnarray}
{1\over {\sigma_0}} {{d\sigma}\over{dx_b}} (x_b,m_Z,m_b) &=&
\sum_i\int_{x_b}^1
{{{dz}\over z}\left[{1\over{\sigma_0}}
{{d\hat\sigma_i}\over {dz}}(z,Q,\mu_R,\mu_F)
\right]^{\overline{\mathrm{MS}}}
D_i^{\overline{\mathrm{MS}}}\left({x_b\over z},\mu_F,m_b \right)} \nonumber \\
&+& {\cal O}\left((m_b/Q)^p\right),
\label{pff}
\end{eqnarray}
where $x_b=2p_b\cdot Q/m_Z^2$ is the bottom energy fraction in the $Z$ rest frame;
$\sigma_0$ is the Born LO cross-section; $\sigma_i$ the cross-section
for the production of a massless parton $i$; $D_i$ the perturbative
fragmentation function expressing the transition of $i$ into a heavy $b$;
$\mu_R$ and $\mu_F$ are the renormalization and factorization scales, respectively, and
$p$ is an integer $p\geq 1$.
Most analyses have so far assumed that,
in  Equation~(\ref{pff}), $i$ is a light quark, which corresponds to the so-called
non-singlet approximation, while  gluon splitting $g\to b\bar b$ is neglected.
This splitting, as well as $g\to c\bar c$ for charm production,
was accounted for in \cite{cno}, but the authors found  
very little impact on the phenomenological results.

The
perturbative fragmentation functions follow the DGLAP evolution equations
and their value at any scale $\mu_F$ can be obtained after an initial condition
at $\mu_{0F}$ is given. The initial condition was first computed at NLO in \cite{mele}
for $e^+e^-$ annihilation into heavy quarks, and then rederived in
\cite{cc}, demonstrating its process independence on more general grounds.
As pointed out above, the NNLO initial conditions for quarks and gluons
can be found in  \cite{melni,alex}, respectively.
Moreover, one can prove, in the same manner as for
parton distribution functions, that solving the DGLAP equations allows one to resum the large heavy-quark mass logarithms, i.e., $\ln(m^2/Q^2)$,
which appear in the massive NLO spectrum (collinear
resummation). In particular, using both NLO
initial condition and splitting functions yields next-to-leading
logarithmic (NLL) collinear resummation, i.e., for an evolution
from $\mu_{0F}\simeq m$ to $\mu_F\simeq Q$, terms $\alpha_S^n\ln^n(m^2/Q^2)$
(LL) and  $\alpha_S^n\ln^{n-1}(m^2/Q^2)$ (NLL) are resummed.
Including NNLO corrections to
the splitting functions \cite{mmv} and the initial condition
\cite{melni,alex} would potentially extend collinear resummation to NNLL.
The DGLAP equation is typically solved in 
Mellin moment space \endnote{The Mellin transform of a function $f(x)$,
  with $0<x<1$, is
  defined as $f_N=\int_0^1{dx x^{N-1} f(x)}$.},
  where convolutions are turned into ordinary products,
and then $x$-space results are recovered after an inverse Mellin transform,
which usually follows the minimal prescription \cite{mini}.

Both the initial condition and coefficient functions of the main heavy-quark
production processes exhibit terms behaving like
$\sim\alpha_S[\ln(1-x_b)/(1-x_b)]_+$
and $\sim\alpha_S/(1-x_b)_+$, which become large for soft or collinear radiation,
i.e., $x_b\to 1$, which corresponds to $N\to\infty$ in moment space.
The authors of Refs.~\cite{cc,cno,ccm,corc}
performed threshold resummation in the NLL approximation in Mellin space, i.e.,
terms $\alpha_S\ln^{n+1}N$ and $\alpha_S^n\ln^nN$ in the Sudakov
exponent are summed to all orders, and inverted the result back to $x$-space.
The authors of Refs.~\cite{agl,corfer}  instead implemented
NNLL large-$x$ resummation in both the initial condition and the coefficient function,
though matching it to NLO fixed-order results for bottom and charm fragmentation
in $e^+e^-$ collisions.

Before discussing hadronization corrections, it has to be pointed out that,
although all such heavy-quark calculations resum threshold contributions with
high accuracy, as discussed, f
or instance, in \cite{cc} at NLL and in \cite{vogt}
at NNLL or NNNLL, they are still not reliable at  very large $x$, since the
Sudakov exponent exhibits a branch point, related to the Landau pole of the
strong coupling constant.
This effect is especially relevant 
in the initial condition, where renormalization and
factorization scales vary around the heavy-quark mass. The branch point is
found when the Mellin variable $N\sim m/\Lambda$ or $x\sim 1-\Lambda/m$, with 
$\Lambda$ being the QCD scale in the
$\overline{\rm MS}$ renormalization scheme.
Due to this issue, one can already envisage that convoluting
parton-level calculations
with simple non-perturbative models or implementing an
effective coupling in some given approximation
will not be enough to obtain reliable predictions
for very high $x$ values.

As a whole, while all perturbative
calculations to extend heavy-quark fragmentation to NNLO+NNLL have been
available for a while in a number of processes, the state of the art
is generally NLO+NLL.
A remarkable study was carried out in \cite{scet}, which calculates 
heavy-quark fragmentation in $e^+e^-$ annihilation in the NNLO approximation,
with NNLL DGLAP evolution and NNNLL threshold resummation, 
within the framework of the soft collinear effective
theory (SCET). Work towards a NNLO+NNLL calculation in the
perturbative fragmentation framework for $e^+e^-$ collisions
is currently underway \cite{prog}.
Beyond NLO+NLL, the work in ~\cite{terry} describes, 
in the top-quark narrow-width approximation, 
$t\bar t$ production and bottom 
fragmentation in top decays at NNLO
in the framework of perturbative fragmentation functions, with NNLL DGLAP
evolution and NNLL threshold resummation in the initial condition, though
with no
large-$x$ resummation in the top-to-bottom coefficient function.

\section{Non-Perturbative Corrections to Heavy-Quark Fragmentation\label{sec3}}

For the sake of describing experimental data on heavy-hadron production,
it is necessary to convolute the perturbative spectrum with
a non-perturbative fragmentation function, which typically contains parameters which are to be tuned to
experimental data. In particular, simple power laws are often used as non-perturbative fragmentation functions and a well-known example is the model proposed in
\cite{kart}, which has one
tunable parameter $\beta$, and $x$ is the heavy-hadron energy fraction
with respect to the fragmenting quark:
\begin{equation}\label{kk}
  D_{np}=(1+\beta)(2+\beta)(1-x)x^\beta.
  \end{equation}
By relying on the universality of the hadronization transition, one typically tunes models such as Equation~(\ref{kk}) to the most precise data available,
such as those from
$e^+e^-$ machines, and then uses the best-fit parameters in other environments, such as hadron
colliders. For the sake of consistency, the same accuracy and perturbative
parameter settings are to be used in calculating both the
perturbative process, which is used  for hadronization tuning, and the one to which the best
fit is applied. However, a drawback of this procedure
is that, as long as one computes the parton-level process to a finite
accuracy, albeit resummed in a given
logarithmic approximation, there are missing corrections which are
process-dependent, which
makes the tuning method and hadronization model not really universal. 
Nevertheless, while any improvement in perturbative calculations as well
as in modelling non-perturbative corrections for processes involving
heavy quarks would be certainly desirable, for the time being, fitting
a non-perturbative fragmentation function to precise
data from $e^+e^-$ experiments,
e.g., at LEP \cite{aleph,opal,delphi} or SLD \cite{sld},  
and consistently applying the results to other processes still represents
the best way to approach heavy-hadron production in the perturbative
fragmentation approach. Within the standard resummation formalism,
this procedure was carried out in ~\cite{cc,cno,scet}
for heavy-quark production in $e^+e^-$ collisions, and in \cite{ccm,corc} for
$B$-hadron production in top ($t\to bW$) and Standard Model Higgs
($H\to b\bar b$)  decays, respectively. 
Strictly speaking, the fits carried out in the literature should be used as long
as the same hadron species are involved; in particular, while OPAL \cite{opal}, DELPHI \cite{delphi}, and SLD \cite{sld} reconstructed both mesons and baryons,
such as the $\Lambda_b$, the ALEPH sample~\cite{aleph} contained only mesons.
As a matter of fact, since the baryon fraction is estimated to be
of the order of $10\%$, $B$-hadron data are often taken together when
tuning non-perturbative models. For example, Refs.~~\cite{drol,mescia},
which compared Monte Carlo event generators with resummed computations for
$b$-quark fragmentation, fitted the hadronization models to all
LEP and SLD data as if they came from one single experiment.
Within the soft collinear effective theory
formalism, the authors of \cite{scet} fit a hadronization model with two free parameters
\cite{neub} to LEP and SLD data, either altogether or discarding the OPAL
ones. This model, along with its best-fit parameters, was then implemented
in the Monte Carlo code developed in \cite{terry} to describe
$t\bar t$ production and decay at NNLO. 

\begin{figure}[H]
 
  \includegraphics[width=10 cm]{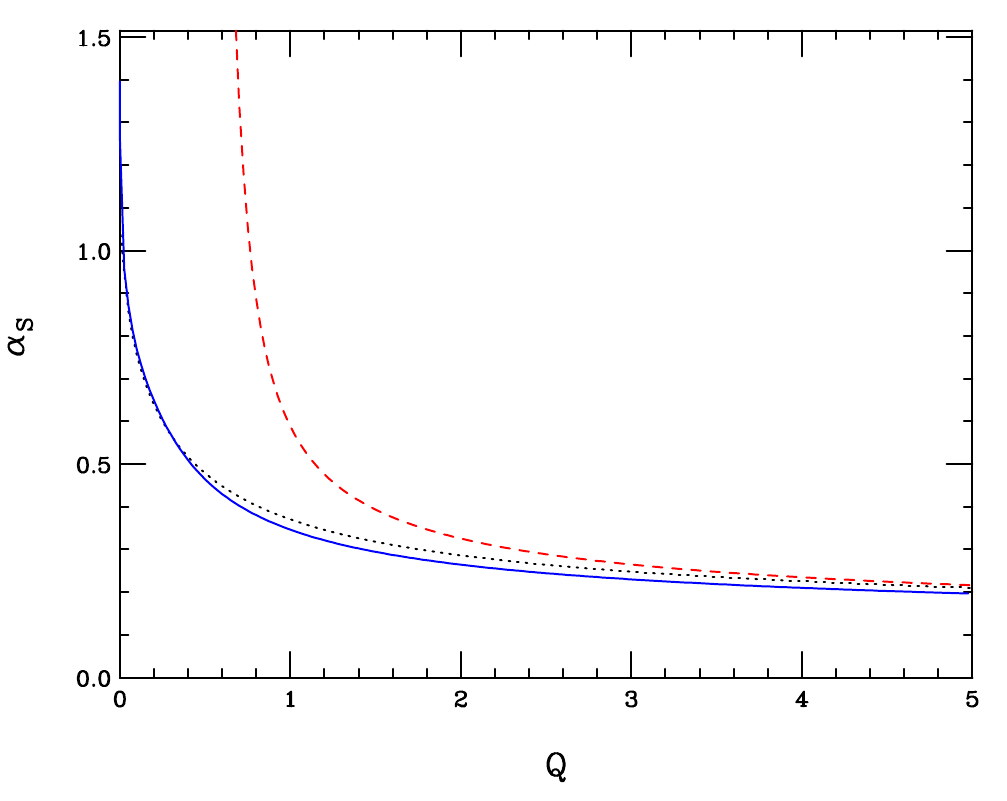}
\caption{Standard (dashes line), analytic (dots), and effective (solid line)
  strong coupling constant at NNLO, as a function of the scale $Q$.}
\label{figas} 
\end{figure}

In the rest of this section, I wish to
review the alternative method, 
based on an effective strong coupling constant, proposed in
\cite{shirkov} and
employed in ~\cite{agl,corfer} for bottom and charm fragmentation, 
respectively.
Above all, the pioneering work in ~\cite{amati} showed that, in
the framework of resummations,
for the sake of summing up subleading soft/collinear contributions,
the momentum-independent
coupling constant $\alpha_S$ is to be replaced by
the following integral over the
discontinuity of the gluon propagator ($1/s$), so that 
the argument of $\alpha_S$ is roughly the transverse momentum of the emitted
parton with respect to the parent one:
\begin{equation}
  \label{alfakt}
\alpha_S\longrightarrow \frac{i}{2 \pi}\int_0^{k_T^2} ds 
\ {\rm Disc}_s\  \frac{\alpha_S(-s)}{s}\simeq \alpha_S(k_T^2)
\ \ \ ;\ \ \ 
      {\rm Disc} \ f(s)=\lim_{\epsilon\to 0}\left[f(s+i\epsilon)-f(s-i\epsilon)\right].
        \end{equation}
It is well known that in resummations, as well as
parton showers, the argument of the strong coupling is the parton
transverse momentum. 
The integral (\ref{alfakt}) is typically performed, neglecting the imaginary part
$\sim i\pi$ in the denominator of $\alpha_S(-s)$, which reads, e.g., at leading
order:
\begin{equation}\label{alfalo}
  \alpha_{S,\rm{LO}}(-s)=\frac{1}{\beta_0 \ln[(-s-i\epsilon)/\Lambda^2]}=
  \frac{1}{\beta_0 \ln[(|s|/\Lambda^2]-i\pi\Theta(s)]},\end{equation}
where $\beta_0$ and $\Lambda$ are the first terms of the beta function and
the QCD scale in the $\overline{\rm{MS}}$ renormalization scheme, respectively,
and $\Theta(x)$ is the Heaviside step function. In other words, the approximate
equality in Equation~(\ref{alfakt}) assumes
$\ln(|s|/\Lambda^2)\gg \pi$ in the denominator 
Equation~(\ref{alfalo}), which is clearly questionable, since
one sums soft and collinear parton radiation to all orders, namely
partons with small virtualities $s$. In fact, as discussed before,
when using the standard coupling, resummed calculations are not fully reliable
for very soft or collinear emissions, and even including
extra non-perturbative
models, such as the power law in
(\ref{kk}), they fail to describe
heavy-hadron energy spectra
for  very large values of $x$.
Following \cite{shirkov}, the work in \cite{agl,corfer} first defines an analytic
coupling which is free from the Landau pole:
\begin{equation}\label{analytic}
 \bar\alpha_S(Q^2)= 
\frac{1}{2\pi i}
\int_0^{\infty}\frac{ds}{s+Q^2} 
{\rm Disc}[\alpha_S(-s)]\end{equation} and then constructs an
effective coupling constant by inserting Equation~(\ref{analytic})
in (\ref{alfakt}):
\begin{equation}\label{alfaeff}
\tilde\alpha_S(k_T^2)=
\frac{i}{2 \pi}\int_0^{k_T^2} ds 
\ {\rm Disc}\frac{ \bar\alpha_S(-s) }{ s }.\end{equation}
At LO, for simplicity, one obtains
\begin{equation}
\bar\alpha_{S,\rm{LO}}(Q^2)=
{1\over\beta_0}\left[{1\over{\ln(Q^2/\Lambda^2)}}-
  {{\Lambda^2}\over{Q^2-\Lambda^2}}\right],\end{equation}
which clearly shows that the integrand function in (\ref{alfaeff}) is 
free from the Landau pole, and
\begin{equation}
\tilde\alpha_{S,\rm{LO}}(Q^2)=
\frac{1}{2\pi i\beta_0}\left[\ln\left(\frac{Q^2}{\Lambda^2}+i\pi\right)
  -\ln\left(\frac{Q^2}{\Lambda^2}-i\pi\right)
  \right].\end{equation}
One can determine the relation between standard
and effective coupling constants, showing that they start to differ
from ${\cal O}(\alpha_S^3)$ as follows:
\begin{equation}
\tilde\alpha_S(Q^2)=\alpha_S(Q^2)-\frac{(\pi\beta_0)^2}{3}\alpha_S^3(Q^2)+
{\cal O}(\alpha_S^4).
\end{equation}
In Figure~\ref{figas}, we present the standard (dashes), analytic (dots),
and effective (solid) coupling constants as a function of the scale $Q$, and
observe that for $Q>5$~GeV, the three couplings agree. At lower
energies, the standard one significantly deviates, to the point of
diverging when $Q$ becomes close to the Landau pole, while the effective and analytic
couplings are close to each other, differ by about 10\% for $Q>0.5$~GeV,
and roughly agree again at very low $Q$.

The assumption adopted in \cite{agl,corfer} states that
it is enough to replace the standard with the effective coupling in the
perturbative calculation, i.e., $\alpha_S(Q)\to \tilde\alpha_S(Q)$,
in such a way that a prediction obtained for a heavy quark,
such as the $b$ quark, 
can be applied to a heavy hadron, i.e., a $B$ meson or baryon.
In the simple formulation adopted in \cite{agl,corfer}, based on
~\cite{shirkov}, there is indeed no distinction between mesons and
baryons, as well as
spin-0 and spin-1 hadrons. Of course, one could add extra parameters
to distinguish between hadron species, but in this way the effective coupling
model would lose its
peculiar feature of being free from non-perturbative parameters.

\textls[-15]{As discussed in \cite{agl,corfer}, the effective coupling constant
defined through Equation~(\ref{alfaeff})} was implemented in the NNLO approximation,
along with a calculation for $e^+e^-$ annihilation which uses NLO
coefficient functions and an initial condition of the perturbative
fragmentation function, NLL DGLAP evolution in the non-singlet
approximation, and NNLL large-$x$ resummation in the initial condition.
Such an accuracy in the perturbative calculation should be
seen as a part of the model and, as explained in \cite{agl}, it was
dictated by the available precision of computations at that time and, above
all, by the fact that it led to an overall reasonable comparison with the data.
Before presenting the results yielded by the model, I wish to stress
that another difference between the method which uses a
non-perturbative fragmentation and the one based on an effective coupling
constant. While in the first case one fixes the perturbative parameters
when tuning the model and only accounts for the uncertainty in the fitted parameters, in the
second one there is no tuning and therefore the dependence of the
prediction on the perturbative entries, such as scales and heavy-quark masses,
is to be estimated.
\vspace{-5pt}
\section{Results---\boldmath{$B$} Production}\label{sec4}
\begin{figure}[H] 
\includegraphics[width=10 cm]{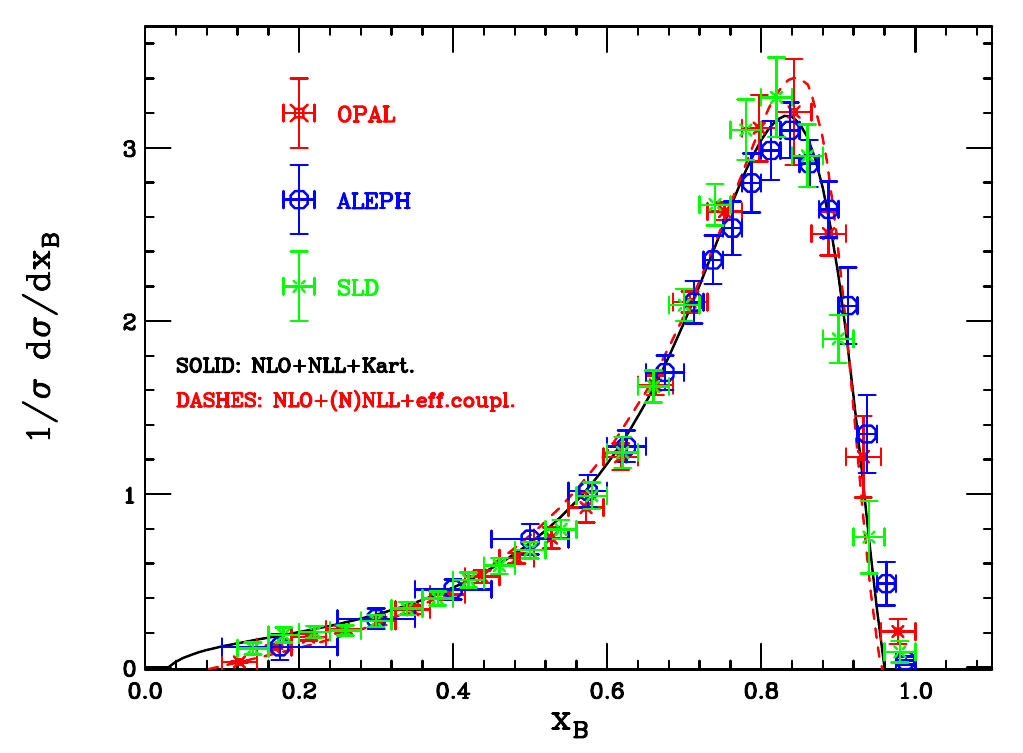}
\caption{Experimental data on $B$-hadron production at SLD, OPAL, and ALEPH,
  compared with theoretical calculations carried out in the
  perturbative fragmentation framework and using either a tuned
  hadronization model (solid) or an effective coupling constant (dashes)
  to model non-perturbative corrections.}
\label{eebb} 
\end{figure}

In this section, I shall present some results on $B$-hadron production
using models such as the one in
(\ref{kk}) or the effective strong coupling constant (\ref{alfaeff}).
The results will be expressed in terms of $x_B$,
namely the $B$-hadron energy fraction in the centre-of-mass frame, which
is the hadron-level counterpart of $x_b$ introduced in Section~\ref{secper}:
\begin{equation}
  x_B=\frac{2p_B\cdot Q}{m_Z^2},
\end{equation}
where $p_B$ is the $B$-hadron four-momentum.
Starting from $e^+e^-$ annihilation into bottom pairs, 
the NLO+NLL calculation in \cite{cc} can be convoluted with the model
(\ref{kk}) and its free parameter $\beta$ tuned to ALEPH \cite{aleph},
OPAL \cite{opal}, and SLD \cite{sld} data.
The authors of \cite{drol,mescia} fitted such data as if they came from one single
experiment in the range $0.18<x_B<0.94$ and
neglected the correlations among data points.
Discarding a few data points at low and large $x_B$ is necessary to obtain
a reasonable fit and is justified by the fact that 
the coefficient function exhibits terms behaving $\sim\ln x_b$ and both the
coefficient function and the initial condition present contributions
$\sim\ln(1-x_b)$, which are enhanced for $x_b\to 0$ and $x_b\to 1$ and have
not been resummed. Furthermore, at very large $x$, one is mostly
sensitive to the non-perturbative regime and resumming threshold
contributions at NLL or NNLL, as well as modelling hadronization effects
via an effective coupling constant or the model (\ref{kk}), is not
enough to be reliable for $x_B\to 1$ or $N\to \infty$ in moment space.
In fact, the authors of \cite{cno} improved the large-$N$ behaviour of the heavy-quark spectra, and consequently the large-$x_b$ one as well, by rescaling the
$N$ variable, according to
\begin{equation}\label{rescale}
 N\to N\frac{1+f/N'}{1+fN/N'},  
\end{equation}
where $N=\exp[1/(b_0\alpha_S(\mu^2)$], with $\mu=\mu_R$ in the coefficient
  function and $\mu=\mu_{0R}$ in the initial condition.
  The reason of the rescaling (\ref{rescale}) is that, as anticipated
  before, 
  the Sudakov exponent, resumming threshold contributions, exhibits
branch points  $\sim \ln(1-\lambda)$, with $\lambda=b_0\alpha_S(\mu^2)\ln N$,
  which are divergent for $\lambda\to 1$, namely for $N\simeq \mu^2/\Lambda$,
  if one uses the LO expression for $\alpha_S(\mu)$. The very fact that the
  divergence involves the QCD scale $\Lambda$, and hence the Landau pole of
  the strong coupling, means that it is unphysical and stresses
  the unreliability of perturbative calculations in this regime.
  If one roughly sets $N\sim 1/(1-x_b)$, one obtains that
  $x_b\simeq 1-\Lambda/\mu$ is the equivalent relation in $x_b$-space where
  the resummed computation can be trusted, i.e.,
roughly  $x_b<0.96$ for bottom production.
  Thanks to the replacement (\ref{rescale}) and fitting
  the parameter $f$, the authors of ~\cite{cno} managed to
  improve the large-$x_b$ behaviour of the spectrum.
  In this paper, however, we follow the approach \cite{cc,agl} and 
  do not perform the replacement (\ref{rescale}), without adding any extra
  parameter, such as $f$.
  By setting the perturbative parameters to values
\begin{equation}
  \mu_R=\mu_F=m_Z,\ \mu_{0R}=\mu_{0F}=m_b,\ m_Z=91.19~{\rm GeV},\
m_b=5~{\rm GeV},\   \Lambda=200~{\rm MeV}, \label{param}\end{equation}
we obtain:
\begin{equation}
  \label{best}
  \beta\simeq 17.18\pm 0.30,
\end{equation}
with $\chi^2/{\rm dof}\simeq 46/53$, from the fit.
The uncertainty on $\beta$ in Equation~(\ref{best}) is only due to the experimental
errors on the data, while no perturbative uncertainty is accounted for.
Regarding the effective coupling model, 
in ~\cite{agl} the authors varied all perturbative
parameters in Equation~(\ref{param}) in typical ranges, i.e., the scales were
allowed to change between half and twice the central values, 
and found a reasonable
overall agreement in the same
range as the model (\ref{kk}), i.e., $0.18<x_B<0.94$.
As for the bottom-quark mass, since the model \cite{shirkov}
assumes that, whenever the effective strong coupling is used, the $b$ quark
behaves as a $B$-hadron, it is not uniquely determined whether $m_b$ should be the quark or hadron mass. Therefore, ~\cite{agl} made
a conservative choice and varied $m_b$ in a range which was
sufficiently large to include the 
estimations of both quark and hadron masses.
Furthermore, ~\cite{agl} found that the parameters
which have the largest impact are $m_b$ itself and $\mu_{F,0}$, and that the 
best overall fit, i.e.,  $\chi^2/{\rm dof}\simeq 103.0/54$, is obtained for 
$m_b=5$~GeV and $\mu_{0F}=m_b/2$. Moreover, the inclusion of NNLL large-$x$
resummation in the coefficient function and the initial condition of the
perturbative fragmentation function turned out to play a major role in the
comparison with the data.
Several comparisons among $B$-hadron data and
calculations of $e^+e^-\to b\bar b$
processes were presented in ~\cite{cc,cno,drol,agl,mescia}.
Figure~\ref{eebb} presents an example of such a comparison, namely the NLO+NLL
calculation with the hadronization model (\ref{kk}) and the NLO+(N)NLL
computation which uses the effective coupling constant confronted
with OPAL, ALEPH and SLD data altogether.
Overall, the agreement of two calculations with the data is acceptable;
the two theoretical predictions roughly agree for large and average $x_B$
values, while the spectra yielded by the model (\ref{kk}) are above
the one relying on the effective coupling at small $x_B$ and
above around the peak.

An alternative method to the fits in $x_B$-space involves working 
in Mellin space, which is feasible as long as the experimental
collaborations provide the moments of the $b\bar b$ cross-section.
As discussed in \cite{cn}, an advantage of this procedure is that one does not have to rely on
any hadronization model and 
perform any inversion from $N$- to $x_B$-space. In fact, the moments of
the non-perturbative fragmentation function can be obtained by dividing the
experimental ones by the theoretical $N$-space cross-section.
As for the effective coupling model, no non-perturbative
fragmentation function is fitted, so that one is provided directly with
the moments of the $B$-hadron
production cross-section, as a function of the entries in the
perturbative computation, which can be directly compared with the data.

As an example of this procedure, in Table~\ref{tabmom}, 
the first
four moments of the $B$-hadron cross-section are presented,
as measured by DELPHI \cite{delmom}, along with the prediction
of ~\cite{agl} with an effective coupling constant,
as well as the NLO+NLL parton-level result
computed in \cite{cc} in $N$-space and the extracted moments of the
non-perturbative fragmentation function.
The prediction of ~\cite{agl} accounts for the errors due to the
variation of the perturbative parameters summed in quadrature and, within
the uncertainties, agrees with the moments in \cite{delmom}.
For the NLO+NLL calculation, the quoted moments are computed using the
parameter values in Equation~(\ref{param}).

\vspace{-6pt}
\vspace{-6pt}
\begin{table}[H]
\caption{\label{tabmom} Experimental Mellin moments
    of the $B$-hadron cross-section from the
    DELPHI collaboration, along with the moments yielded by
    the calculation in \cite{agl}, based on an effective coupling constant,
    the perturbative moments at NLO+NLL as in ~\cite{cc}, and
    the extracted $N$-space non-perturbative fragmentation function.} 
\setlength{\cellWidtha}{\textwidth/5-2\tabcolsep+0in} 																
\setlength{\cellWidthb}{\textwidth/5-2\tabcolsep-0in}																	
\setlength{\cellWidthc}{\textwidth/5-2\tabcolsep-0in}	
\setlength{\cellWidthd}{\textwidth/5-2\tabcolsep-0in}
\setlength{\cellWidthe}{\textwidth/5-2\tabcolsep-0in}																		
\scalebox{1}[1]{\begin{tabularx}{\textwidth}{>{\centering\arraybackslash}m{\cellWidtha}>{\centering\arraybackslash}m{\cellWidthb}>{\centering\arraybackslash}m{\cellWidthc}>{\centering\arraybackslash}m{\cellWidthd}>{\centering\arraybackslash}m{\cellWidthe}}																	

\toprule
&\boldmath{ $\langle x\rangle$} &\boldmath{ $\langle x^2\rangle$ }&\boldmath{ $\langle x^3\rangle$}
& \boldmath{$\langle x^4\rangle$ }\\
\midrule
$e^+e^-$ data $\sigma_N^B$ & $0.7153 \pm 0.0052$ & 
$0.5401 \pm 0.0064$ &
$0.4236 \pm 0.0065$ & $0.3406 \pm 0.0064$  \\
\midrule 
$[\sigma_N^B]_{\tilde\alpha_S}$  & 
$0.6867 \pm 0.0403$ & $0.5019 \pm 0.0472$ & 
$0.3815 \pm 0.0465  $ & 
$0.2976 \pm 0.0462 $ \\
\midrule
$[\sigma_N^b]_{\rm NLO+NLL}$  & 0.7801 & 0.6436
& 0.5479 & 0.4755\\
\midrule$D_N^B$ &
0.9169 & 0.8392 & 0.7731 & 0.7163\\
\bottomrule
\end{tabularx}}
\end{table}

\begin{figure}[H] 
\includegraphics[width=10 cm]{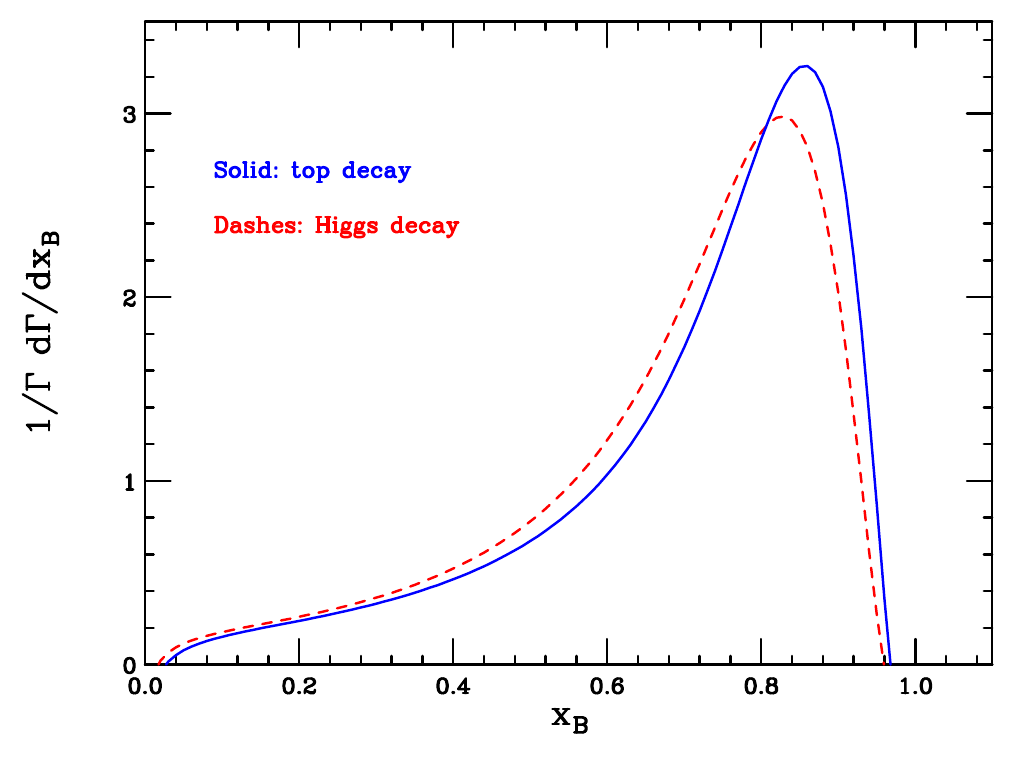}
\caption{$B$-hadron spectrum in top (solid) and Higgs (dashes) decays,
  according to a NLO+NLL perturvative calculation and modelling non-perturbative
corrections by means of the model (\ref{kk}), tuned to $e^+e^-$ data.}
\label{ht}
 \end{figure}

Regarding $B$-hadron production in top decays, as discussed before, the
NLO+NLL $b$-quark spectrum was computed in \cite{cm,ccm}, while
~\cite{corc} performed a companion calculation for
standard model Higgs decay $H\to b\bar b$. Both computations were
undertaken in the narrow-width approximation for top quarks and Higgs
bosons, and convoluted with a non-perturbative
fragmentation such as (\ref{kk}), tuned to $e^+e^-$ data from LEP and
SLD, as discussed above.
Although it would be a straightforward extension, the
model based on the effective strong coupling constant
has not been yet applied to $t\to b W$ and $H\to b\bar b$ processes.
In Figure~\ref{ht}, the $x_B$ spectrum in top and Higgs decays
is displayed, namely the $B$-hadron energy distribution in the top or Higgs
rest frame. The difference is mostly due to the coefficient functions, which, unlike the fragmentation function, are process-dependent, with
$b$-flavoured hadrons harder in top decays with respect to
$H\to b\bar b$ events.
In the narrow-width approximation, the $B$-hadron
spectra in the top and Higgs decays 
in Figure~\ref{ht} are independent of the production process; hence, they 
are roughly valid for both lepton and hadron colliders, regardless of
the centre-of-mass energy.
However, the
comparison with experimental data is not straightforward, since, unlike
$e^+e^-$ annihilation at the $Z$ pole, the top and Higgs rest frames are not
experimentally accessible. 
While calculations such as those in ~\cite{ccm,corc} are quite inclusive and
limited to the prediction of the $x_B$ distributions, 
as discussed in the introduction, ~\cite{terry,prog} provide a
Monte Carlo code for top production and decay, in such a way that one is
able to study other observables. At the moment, ~\cite{terry}
included non-perturbative corrections, as those obtained the SCET analysis in
~\cite{scet}, while ~\cite{prog} aims at a self-consistent
tuning of non-perturbative fragmentation functions
in the NNLO approximation for top decays with NNLL DGLAP evolution and
possibly NNLL large-$x$ resummation
in the initial condition of the perturbative fragmentation function.
Unlike \cite{cm,ccm}, which worked out the evolution equations
in the non-singlet-approximation, ~\cite{terry,prog} also include the
singlet contribution, namely gluon splitting.
In particular, the work in \cite{prog} will provide the experimental community
with a consistent set of NNLO non-perturbative fragmentation functions,
which one will be able to use even in different contexts.

\section{Results---\boldmath{$D$} Production\label{sec5}}

In this section, we shall present a few results on charm-quark fragmentation
and $D$-hadron production
in the perturbative fragmentation framework.
In fact, while the formalism is the same as for bottom-quark and $B$-hadron
production, the scales involved are pretty different; therefore, a
separate investigation is mandatory.
Moreover, as far as $e^+e^-$ annihilation is concerned, we have
precise data both at $B$-factories taken by the CLEO \cite{cleo}
and BELLE \cite{belle} collaborations, as well
at the $Z$ pole, such as the ALEPH data in
~\cite{alephd}.

A thorough analysis of $D$-meson production was carried out in \cite{cno}
at NLO+NLL, rescaling the Mellin variable $N$, as in (\ref{rescale}), and
obtaining good agreement with the data \cite{cleo,belle,alephd}.
In this paper, I highlight
the results in \cite{corfer}, where charmed mesons
were investigated using a NLO perturbative
calculation with NLL DGLAP evolution and NNLL threshold resummation,  
and modelling non-perturbative corrections via an effective
coupling in the NNLO approximation, as in \cite{agl}.
All perturbative parameters were varied in
suitable ranges around the central values in Eq.~(\ref{param}),
but with the $b$-quark mass replaced by the charm one.

\begin{figure}[H] 
\includegraphics[width=10 cm]{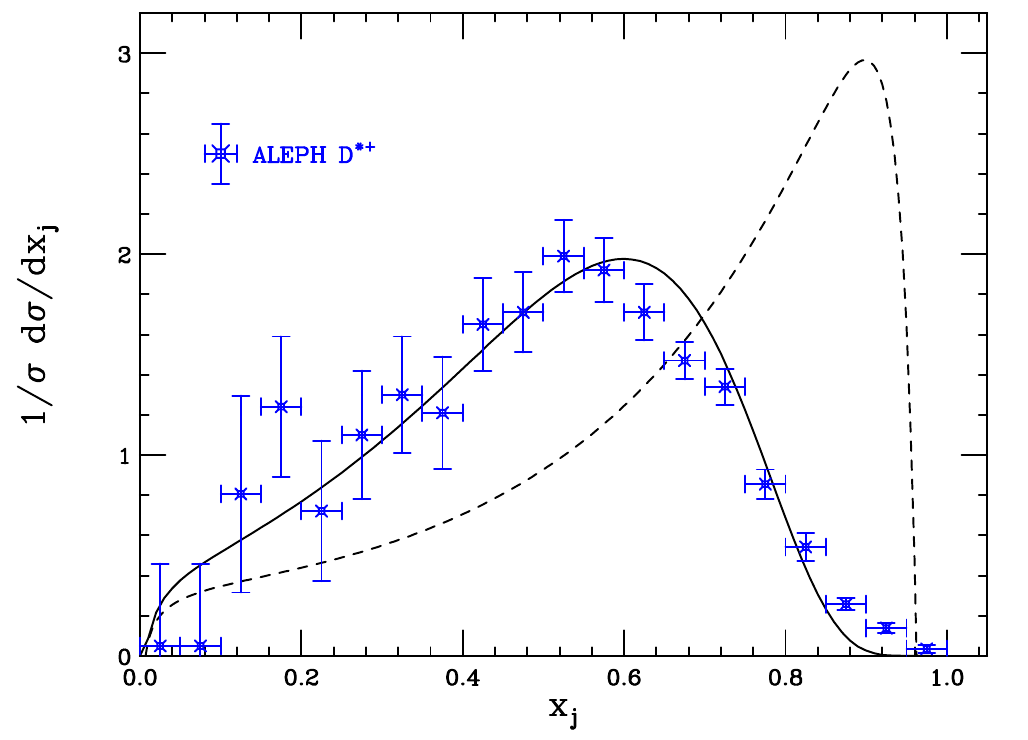}
\caption{Experimental data on $D^{*+}$ production at ALEPH,
  compared with a calculation in the NLO+(N)NLL approximation, as
in \cite{corfer},  using an effective coupling constant (solid)
and $\mu_{0F}=2m_c$ and $m_c=1.8$~GeV in the computation,
with the other parameters set to the values described in the text.
The dashed line corresponds to a parton-level prediction using the standard
strong coupling constant and the same set of perturbative parameters.}
\label{dstar} 
\end{figure}

The work in ~\cite{corfer} presents several predictions for
$D$-meson production in $e^+e^-$ annihilation with the effective coupling
constant, varying the parameters in the perturbative calculation; 
hereafter, I present the results which yield the best comparison with the
experimental data.
In Figure~\ref{dstar}, the ALEPH data \cite{alephd} on $D^{*+}$ production at
LEP are shown, along with the purely perturbative calculation (dashed line)
and the effective coupling prediction obtained 
setting $\mu_{0F}=2m_c$ and $m_c=1.8$~GeV in the parton-level computation, which
means a value of $m_c$ close to the charmed-hadron mass.
Figure~\ref{dstar} displays the strong impact of non-perturbative corrections
and proves that the calculation relying on the effective coupling constant
is capable of giving a good description of the data for $x_D<0.85$
($\chi^2/{\rm dof}\simeq 27.18/17$, neglecting the correlations).

\vspace{-6pt}
\begin{figure}[H] 
  \includegraphics[width=10 cm]{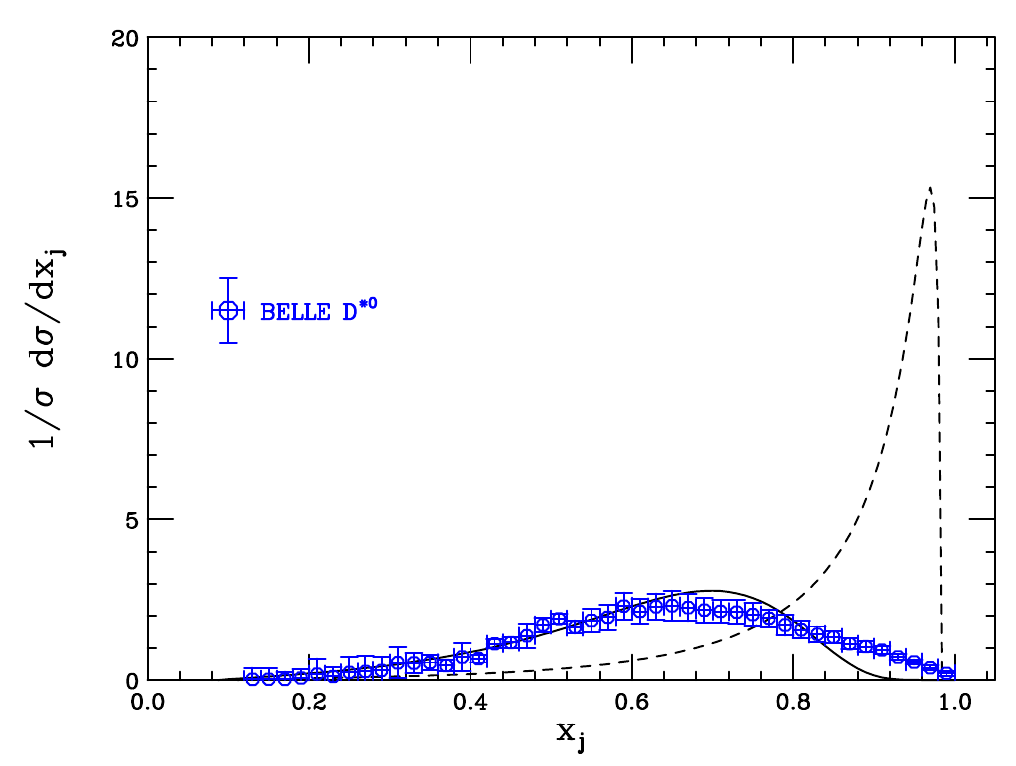}
\caption{BELLE data on $D^{*0}$ production at the $\Upsilon(4S)$ pole,
  compared with the perturbative calculation
  described in the text and in \cite{corfer}, using an effective (solid)
  and standard coupling constant (dashes). Perturbative parameters are set as in the results
  shown
  in Figure~\ref{dstar}.}
\label{beld} 
\end{figure}

As for the analysis at $B$-factories, Figure~\ref{beld} presents the
BELLE data on $D^{*0}$ production at the $\Upsilon(4S)$ pole \cite{belle},
with the best
effective-coupling based prediction (solid line), setting
$\mu_{0F}=2m_c$ and $m_c=1.8$~GeV, and the parton-level result (dashes).
For $x_D<0.85$, using the effective coupling constant, one obtains
$\chi^2/{\rm dof}\simeq 32.10/36$, neglecting the correlations between
data points.
As a whole, one
can learn that at $B$-factories, the partonic spectrum exhibits an even
sharper peak at large $x$ with respect to LEP energies, which is due to the
fact that at the  $\Upsilon(4S)$ pole charm quarks are almost at rest and
have very little phase space to radiate, so that they are mostly produced
at $x\simeq 1$. The impact of the use of the effective coupling is
extremely important, and fundamental to obtain a reasonable
description of the data, after discarding a few data points at very large
energies.

\begin{figure}[H] 
  \includegraphics[width=10 cm]{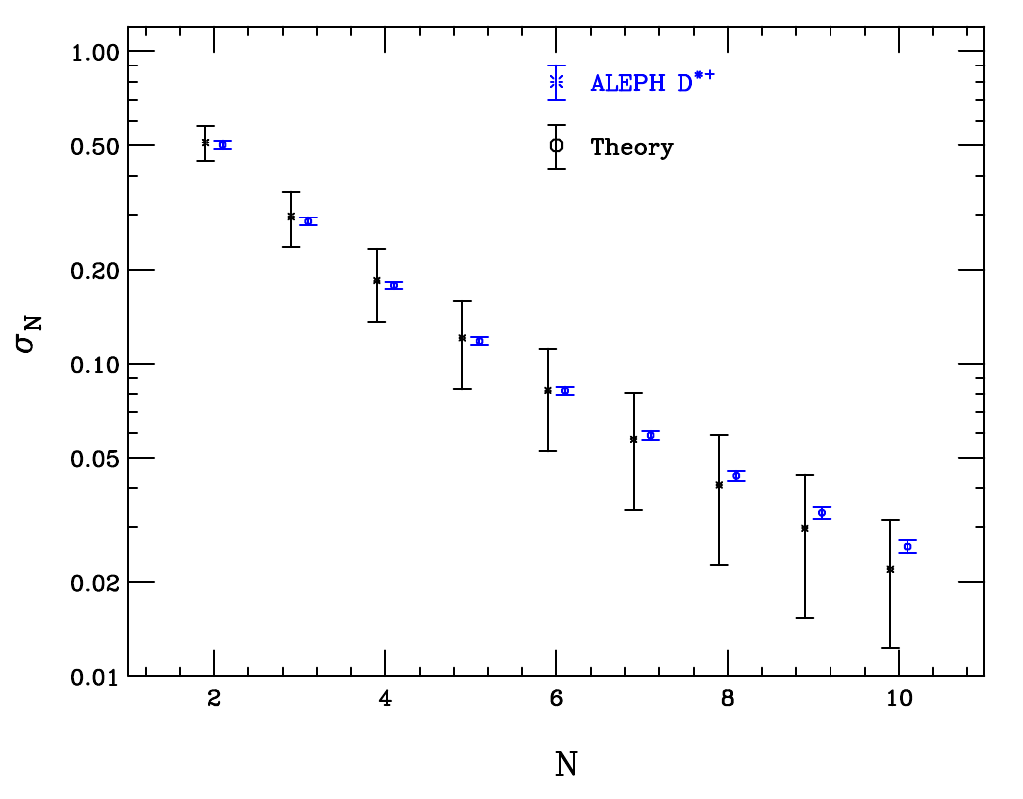}
  \caption{Charmed-hadron production cross-section in $N$-space according
    to the effective coupling model (denoted by `Theory') described in the
    text, 
    compared with the ALEPH moments for $D^{*+}$ production.
    The theory errors are computed as discussed in \cite{corfer}.}
\label{ald} 
\end{figure}

\begin{figure}[H] 
  \includegraphics[width=10 cm]{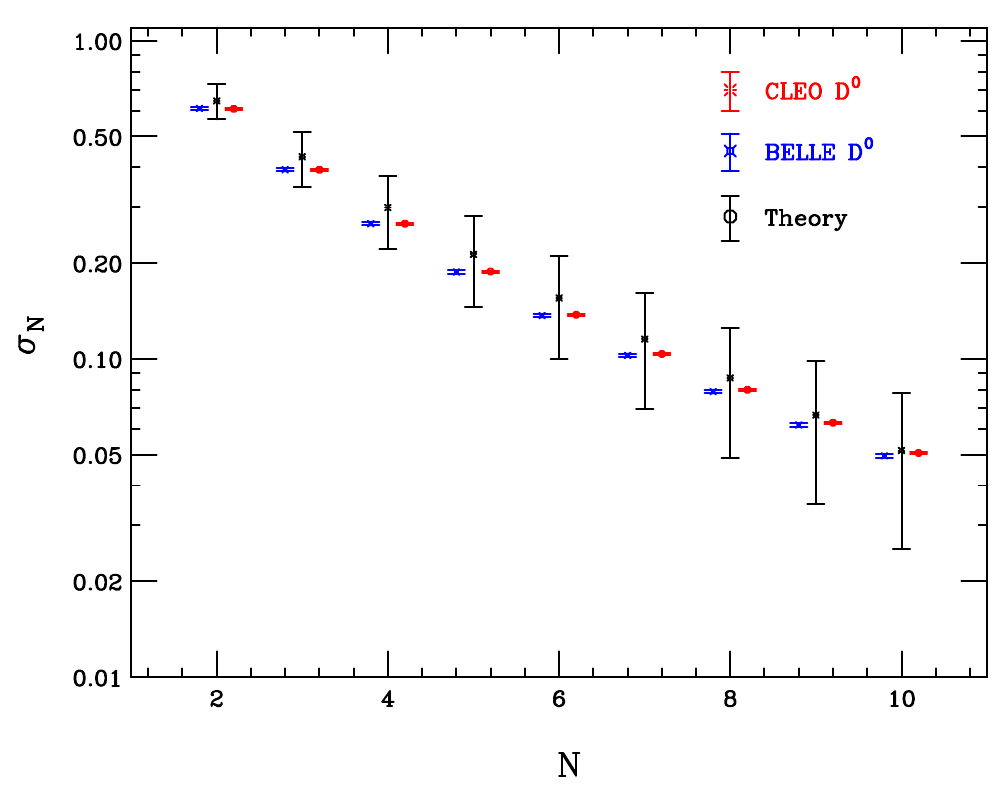}
  \caption{As in Figure~\ref{ald}, but comparing the effective coupling
    prediction with the BELLE and CLEO moments on $D^0$ production.}
\label{bfd0} 
\end{figure}

\begin{figure}[H] 
  \includegraphics[width=10 cm]{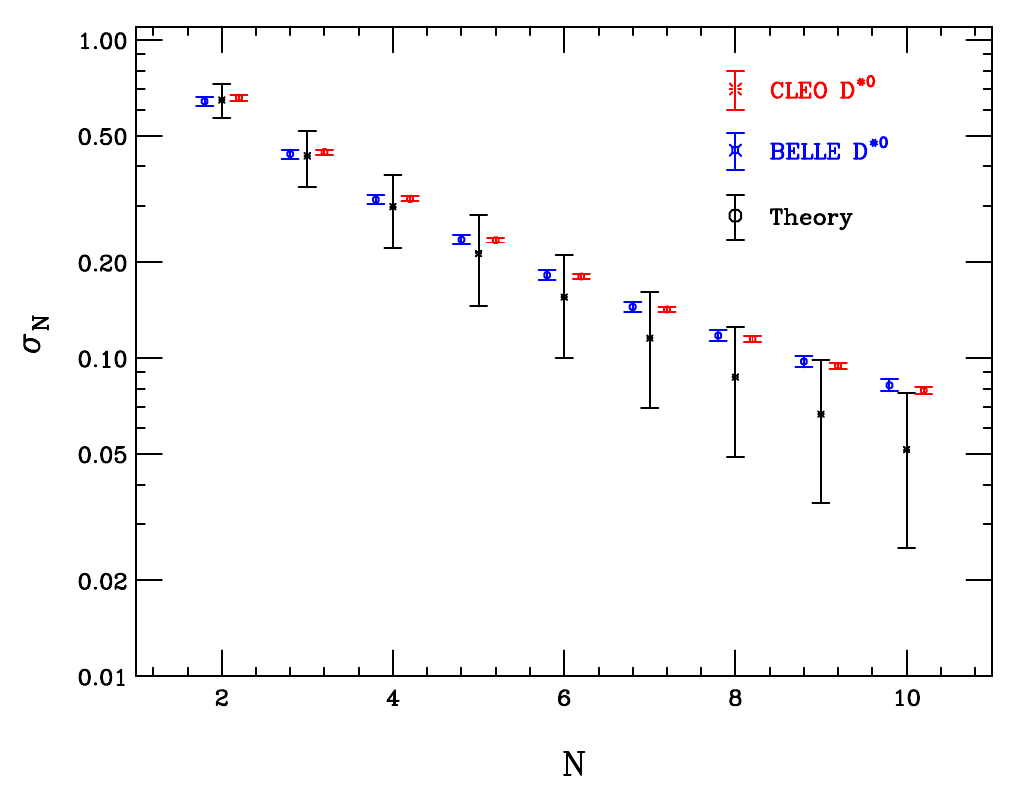}
  \caption{As in Figures~\ref{ald} and \ref{bfd0},
    but with the theory moments faring
    against BELLE and CLEO data on $D^{*+}$ production.}
\label{bfds} 
\end{figure}

Still on the comparison with $B$-factories,
~\cite{corfer} shows that serious discrepancies are
instead present when comparing with CLEO data on $D^0$, $D^{*0}$, and $D^{*+}$
production and with the $D^{*+}$ spectrum measured by BELLE as well,
which clearly indicates that the effective coupling model has to be improved,
especially in the charm sector.
Nevertheless, one finds that, within the errors, one is able of reproducing
the Mellin moments of all considered data sets, as can be noticed in
Figures~\ref{ald}--\ref{bfds}.
This result is remarkable, especially for the
comparison with the BELLE and CLEO $D^0$ samples or the CLEO $D^{*+}$
data, which was unsatisfactory in $x$-space.
In fact, when computing the integrals for the Mellin moments,
we have a compensation between regions where the theory prediction
is higher, i.e., at middle values of $x_D$ or around the peak, or below the data,
i.e., at very large $x_D$.
Furthermore, 
as thoroughly discussed in \cite{corfer}, the rough agreement in
Figures~\ref{ald}--\ref{bfds} is clearly biased by the very large theory uncertainties and
only a NNLO+NNLL investigation on $c\bar c$ production, possibly with
the effective coupling constant as well,
could decrease the theoretical errors and
shed light on the comparison with the experimental moments.
We just point out that a similar result was obtained in \cite{drol}
for the purpose of the HERWIG \cite{herwig} generator, which was not
capable of reproducing the $B$-spectrum data in $x$-space, while its moments
agreed with the experimental ones.

\vspace{-6pt}
\vspace{-6pt}

\section{Conclusions}
I presented an overview of a few selected results on heavy-quark
fragmentation, mostly based on the perturbative fragmentation
approach, which allows one to resum large logarithms of collinear
origin. The investigation mainly dealt with bottom and charm production in
$e^+e^-$ annihilation and the possibility of including non-perturbative
corrections through either a parameter-dependent hadronization model or an
effective coupling constant, constructed in a given approximation.

Overall, I found that, using either a tuned hadronization model or an
effective coupling, the data could be described with an acceptable precision
in both $x$- and $N$-spaces. However, as far as the $x$-space comparison is
concerned, unless one wishes to add extra tunable parameters, 
one has to discard data at very small or large $x$ to achieve a reasonable fit.
In fact, the perturbative calculation exhibits either terms which have not been
resummed or contributions for which a NLL or NNLL resummation is
not sufficient to obtain a reliable prediction.
An interesting feature of the comparison with charm-production data
is that the prediction based on the effective coupling constant is not
completely satisfactory in $x$-space, while it fares quite well against
all experimental Mellin moments, which was interpreted as due to the compensation
of discrepancies in different phase-space regions
when performing $x$-space integrals.

In summary, this paper highlights a few relevant results in the field
of heavy-quark fragmentation, but also stresses some points which deserve
further investigation and extension. Above all, all ingredients to extend
the parton-level calculation from NLO+NLL to NNLO+NNLL, where the logarithmic
accuracy refers to both DGLAP evolution and threshold resummation, are
available and should yield results pretty stable and reliable.
In particular, it will be very interesting to understand how the
$x\to 1$ part of the spectrum behaves at NNLO+NNLL and whether one still needs
to perform the rescaling in (\ref{rescale}) or discard a few
data points to agree with the experimental measurements.
Furthermore, there are small-$x$ logarithms which affect the low-energy
spectra and need to be resummed to have a sensitive prediction for
$x\to 0$ too. Regarding the inclusion of non-perturbative power corrections, one can certainly use the model (\ref{kk}) or the effective strong coupling
constant in conjunction with a NNLO+NNLL perturbative calculation.
Moreover, one can even think of adding more parameters to the non-perturbative
fragmentation function, as done, for example, in the context of
parton distributions, ideally aiming at $\chi^2/{\rm dof}\to 0$. 
Such an approach will be followed in the work in progress on top-pair production
and decay at NNLO+NNLL in ~\cite{prog}.

\acknowledgments{The presented results were obtained in collaboration with
  Ugo Aglietti, Matteo Cacciari, Volker Drollinger, Giancarlo Ferrera,
Federico Mescia, and Alexander Mitov.}


\begin{adjustwidth}{-\extralength}{0cm}
 \printendnotes[custom] 

\reftitle{References}

\end{adjustwidth}
\end{document}